# Spin-Phonon Coupling and Thermodynamic Behaviour in YCrO$_3$ and LaCrO$_3$: Inelastic Neutron Scattering and Lattice Dynamics

Mayanak K. Gupta[1$], Ranjan Mittal[1,2*], Sanjay K. Mishra[1], Prabhatasree Goel[1], Baltej Singh[1,2], Stephane Rols[3] and Samrath L. Chaplot[1,2]

[1]*Solid State Physics Division, Bhabha Atomic Research Centre, Mumbai, 400085, India*
[2]*Homi Bhabha National Institute, Anushaktinagar, Mumbai 400094, India*
[3]*Institut Laue-Langevin, 71 avenue des Martyrs, Grenoble Cedex 9, 38042, France*
Corresponding Authors Email: mayankg@barc.gov.in[$], rmittal@barc.gov.in[*]

We report detailed temperature-dependent inelastic neutron scattering and ab-initio lattice dynamics investigation of magnetic perovskites YCrO$_3$ and LaCrO$_3$. The magnetic neutron scattering from the Cr ions exhibits significant changes with temperature and dominates at low momentum transfer regime. Ab-inito calculations performed including magnetic interactions show that the effect of magnetic interaction is very signicant on the low- as well as high-energy phonon modes. We have also shown that the inelastic neutron spectrum of YCrO$_3$ mimics the magnon spectrum from a G-type antiferromagnetic system, which is consistent with previously reported magnetic structure in the compound. The ab-initio lattice dynamics calculations in both the compounds exhibit anisotropic thermal expansion behaviour in the orthorhombic structure and predict negative thermal expansion along the crystallographic a-axis at low temperatures. We identify the anharmonic phonon modes responsible for this anamolous behaviour in LaCrO$_3$ involving low-energy La vibrations and distortions of the CrO$_6$ octahedra.

## I. Introduction

Materials with more than two switchable ferroic properties are of great interest due to potential use in four state memory devices. Materials that exhibit spontaneous ordering of magnetization, electric polarization and elastic strain are called multiferroics[1-7]. These materials find several technological applications such as refractory electrodes, thermistors, and thermoelectric materials. The coupling of the ferroelectric and ferromagnetic order parameters is particularly attractive if found in the same phase, both from fundamental science and technology point of view[8]. These compounds find use in spintronics and as data storage devices. It is interesting to note that most of the multiferroics, currently of interest are perovskite ($ABO_3$) structured. These structures have a capacity to accommodate multitude of structural distortions and are able to incorporate almost every element. Ideal perovskite structure is a framework of corner sharing octrahedra hosting B cations, while A cations are placed in the resulting dodecahedral sites. Structural instabilities of these biferroic oxides ($ABO_3$) depend on the A cation size (with same B cation) and change in B cation. The effects of external parameters like temperature, pressure, chemical composition drive these distortions, which lead to several rich and unique physical properties. These physical properties are derived from tilt of $BO_6$ octahedra, polar cation displacement leading to ferroelectricity, etc. Multiferroic materials with coupled Ferromagnetic (FM) and Ferroelectric (FE) order parameters are promising for developing new generation of electrically and magnetically controlled multifunctional devices. A large number of studies on chromate based perovskites have been performed in recent years due to its rich physics and simple structures[7,9-12].

There are two types of multiferroics: i) The materials that exhibit magnetic ordering and ferroelectric ordering up to a very high temperature, but the ordering temperature as well as origin of ferroic properties are very different, hence coupling is very weak; ii) On the other hand, there are compounds with the same ordering temperature and origin of ferroelectric and magnetic ordering with strong coupling, but unfortunately the transition temperature as well as the magnitude of electric dipole moments are very small[13,14]. Hence, across the globe intense research is going on to achieve both the requirement of strong coupling as well as high transition temperature.

The compounds $RCrO_3$ (R=Y, La) are magnetic but do not exhibit ferroelectricity in pure single phase. However, mixing them together results in enhanced ferroelectric properties[15-23]. The mixing leads to strain, which could cause the strain mediated ferroelectricity in the mixed phase. Thermal expansion could also act as a source of strain in the mixed compound. Hence, it is important to understand the thermal expansion behavior and its anisotropy that would be useful to design the multiferroic material in future.

RCrO$_3$ (R=Y, La) are found in orthorhombic structure (space group Pnma) at ambient condition (**Fig 1**). With respect to the ideal cubic perovskite structure (space group Pm-3m), this orthorhombic phase is obtained by an anti-phase tilt of adjacent CrO$_6$ octahedra; the R cations is usually in 8 coordination (**Fig 1**). Although Pbnm is a centrosymmetric space-group and non-compatible to ferroelectricity, but local non-centrosymmetric nano regions give rise to ferroelectricity in YCrO$_3$ [18,23-26]. Previous work on YCrO$_3$ indicated[27] that it is ferrimagnet with weak ferromagnetism ($T_N$~140 K). Subsequently it has been shown that YCrO$_3$ exhibits[28] canted antiferromagnetic behaviour below 140 K and a ferroelectric transition around 473 K. More recent X-ray and neutron powder diffraction, and Mössbauer spectroscopy measurements on YCrO$_3$ also indicate[29] that it is antiferromagnetic and the direction of the moment on Cr$^{3+}$ ion is along the c-axis. LaCrO$_3$ is also reported[29,30] to be G-type antiferromagnetic, and the direction of the magnetic moment is found along[30] the crystallographic a-axis.

Ferroelectricity is absent in LaCrO$_3$ due to the large size of La$^{3+}$ ion which might prevent some structural instability necessary for ferroelectric transition (in comparison to Y$^{3+}$). In order to understand the structural instabilities in these oxides, several theoretical and experimental studies have been carried out. Understanding the driving force behind ferroelectric instability in oxides like LaCrO$_3$ has warranted several electronic structure studies, and experimental Raman studies[8,19-22,31,32]. Understanding the correlation between individual phonon modes and structural distortions, make interesting ongoing studies. Temperature dependence of Raman measurements[33] on YCrO$_3$ show that the A$_{1g}$ and B$_{2g}$ Raman active modes about ~ 560 cm$^{-1}$ undergo significant changes across the magnetic transition temperature. Therefore, these modes are likely to show strong coupling with magnetic ordering. There have been studies to understand the changes in the physical properties of YCrO$_3$ with doping; it is found that its electrical conductivity increased with Ca-doping. High pressure synchrotron powder diffraction experiments have been carried on YCrO$_3$ up to 60 GPa to study its evolution with increasing pressure[34]. High resolution neutron diffraction studies have been reported to understand the average and local structure in YCrO$_3$ in order to explain the electric polarization seen above 430 K despite the centrosymmetric phase[18].

In this paper, we report neutron inelastic scattering measurements over a wide range of momentum transfers, which enables to separately identify the contributions from magnetic excitations (at low momentum transfers) and phonons (at high momentum transfers). The neutron measurements have been performed as a function of temperature, which reveal significant changes across the magnetic phase transition; however, the magnetic scattering persists in the paramagnetic phase due to paramagnetic relaxations. We have also performed extensive ab-initio calculations of the phonon spectra and identified the effect of magnetism. Further, we use the ab-initio calculations to derive the anisotropic elasticity and the thermal expansion, including the negative thermal expansion along the a-axis of the orthorhombic structure. The results are in very good agreement with the available data. The theory helps to identify the anharmonic phonon modes behind the anomalous thermal expansion behaviour.

## II. Experimental Details

Polycrystalline samples of $LaCrO_3$ and $YCrO_3$ were prepared by solid state reaction method. The inelastic neutron scattering measurements were carried out using the high-flux time-of-flight (IN4C) spectrometer at the Institut Laue Langevin (ILL), France, covering a wide range of scattering angles from 10° to 110°. Thermal neutrons of wavelength 2.4 Å (14.2 meV) are used for the measurements. The scattering function S(Q, E) is measured in the neutron energy gain mode with a momentum transfer, Q, extending up to 7 Å$^{-1}$. About 10 grams of polycrystalline sample has been used for the measurements over 100-550 K. The polycrystalline sample was put inside a cylindrical niobium sample holder and mounted in a cryoloop. The data analysis was performed by averaging the data collected over the angular range of scattering using ILL software tools[35] to get neutron cross-section weighted phonon densities of states. The phonon density of states $g^{(n)}(E)$ in the incoherent[36,37] one-phonon approximation is extracted from the measured dynamical structure factor S(Q,E) as follows

$$g^{(n)}(E) = A < \frac{e^{2W(Q)}}{Q^2} \frac{E}{n(E,T) + \frac{1}{2} \pm \frac{1}{2}} S(Q,E) > \qquad (1)$$

$$g^{(n)}(E) = B \sum_k \{\frac{4\pi b_k^2}{m_k}\} g_k(E) \qquad (2)$$

Where $n(E,T) = [\exp(E/k_B T) - 1]^{-1}$, ± represents energy gained/lost by the neutron. $b_k$, $m_k$ and $g_k(E)$ are, respectively, the neutron scattering length, mass, and partial density of states of the $k^{th}$ atom in the unit cell. 'A' and 'B' are normalization constants and 2W(Q) is the Debye-Waller factor. The weighting factors $4\pi b_k^2/m_k$ for various atoms in the units of barns/amu are: 0.0866, 0.0695, 0.0671 and 0.2645 for Y, La, Cr and O respectively. The values of neutron scattering lengths for various atoms can be found from Ref.[38]

## III. COMPUTATIONAL DETAILS

The Vienna based ab-initio simulation package (VASP) is used for the calculations[39,40] of the structure and dynamics. The calculations are performed using the projected augmented wave (PAW) formalism of the Kohn-Sham density functional theory within generalized gradient approximation (GGA) for exchange correlation following the parameterization by Perdew, Burke and Ernzerhof.[41,42]. The plane wave pseudo-potential with a plane wave kinetic energy cutoff of 900 eV was adopted. The integration over the Brillouin zone is sampled using a k-point grid of 2×2×2, generated automatically using the Monkhorst-Pack method[43]. The above parameters were found to be enough to obtain a total energy convergence of less than 0.1 meV for the fully relaxed (lattice constants & atomic positions) geometries. The total energy is minimized with respect to structural parameters. The Hellman-Feynman forces are calculated by the finite displacement method. The total energies and force calculations are performed for the 17 distinct atomic configurations resulting from symmetrical displacements of the inequivalent atoms along the three Cartesian directions (±x, ±y and ±z). The convergence criteria for the total energy and ionic forces were set to $10^{-8}$eV and $10^{-5}$eV A$^{-1}$, respectively. The phonon energies (and the dispersion curves and density of states) were extracted from subsequent calculations using the phonopy-1.14 software[44]. The phonon calculation has been done considering the crystal acoustic sum rule. The calculated structures (Table I) of LaCrO$_3$ and YCrO$_3$ agree well with the available experimental data.

The thermodynamic and transport properties of material with magnetic ions also contributed significantly by magnetic interaction between these ions. These additional magnetic interactions between controls translational and rotational flexibility of atoms in the crystal, hence, control the dynamics of the material. In addition to that the associated spins of the magnetic ions form collective spin excitation known as magnons in magnetic ordered phase. While in the paramagnetic phase they do contribute in the free energy in the form of spin configurational entropy. Hence to understand the effect of magnetic interaction on thermodynamics properties, it is important to include the magnetic interaction in the system. The standard DFT formalism allow the inclusion of magnetic interaction in two ways, the collinear spin polarized magnetic calculation where the spin is considered as a one-dimensional variable and the Hamiltonian in rotationally invariant with respect to lattice. The other way to include the magnetic interaction is known as noncollinear spin polarized calculation, this is very computationally expensive method and the spin is treated as a three-dimensional object.

Both the compounds YCrO$_3$ and LaCrO$_3$ are reported[29,30] to be G-type antiferromagnetic. The structure relaxation for both the compounds was performed in the non-magnetic as well as magnetic configurations. For LaCrO$_3$, on-site Hubbard correction is applied within the Dudarev approach[45] using $U_{eff} = U - J = 7.12$ eV[46]. The same value of $U_{eff}$ has been used in the calculations of YCrO$_3$. For YCrO$_3$, we have performed the phonon calculation considering the G-type magnetic configuration and moment on Cr atoms along a-axis (non-collinear calculation) as well along c-axis (collinear calculation). As discussed below, the calculations show that the change in magnetic moment direction does not result in significant changes in the overall phonon density of states. However, we have seen that for some zone centre modes their energy changes significantly (~5%) due to change in the magnetic moment direction. For LaCrO$_3$, we have performed only the collinear calculation. The non-collinear calculation does not converge for required accuracy for phonon calculation, hence we have not performed the non-collinear phonon calculation for LaCrO$_3$.

Thermal expansion behavior of any material is an important thermodynamic property essential for material design for specific application. Here, we have computed the linear thermal expansion behavior in both the compounds using quasi-harmonic approximation that has been found useful in many similar compounds. The thermal expansion calculation is done using pressure dependence of phonon frequencies in the entire Brillouin zone[47]. Each phonon mode of energy $E_{qj}$ ($j^{th}$ phonon mode at wavevector **q** in the Brillouin zone) contributes to the thermal expansion coefficient, which is given by the following relation for an orthorhombic system[48]:

$$\alpha_l(T) = \frac{1}{V_0} \sum_{q,j} C_V(q,j,T) [s_{l1}\Gamma_a + s_{l2}\Gamma_b + s_{l3}\Gamma_c] \ , \ l, l' = a, b, c \ \& \ l \neq l' \quad (3)$$

Where $s_{ij}$ are elements of the elastic compliance matrix, s=C$^{-1}$ at 0 K (C is the elastic constant matrix), V$_0$ is volume at 0 K and $C_V$(**q**, $j$, T) is the specific heat at constant volume due to $j^{th}$ phonon mode at point **q** in the Brillouin zone. The mode Grüneisen parameter of phonon energy $E_{q,j}$ is given as[49],

$$\Gamma_l(E_{q,j}) = -\left(\frac{\partial lnE_{q,j}}{\partial lnl}\right)_{l'} ; \ l, l' = a, b, c \ \& \ l \neq l' \quad (4)$$

The volume thermal expansion coefficient for an orthorhombic system is given by:

$$\alpha_V = (\alpha_a + \alpha_b + \alpha_c) \quad (5)$$

The elastic constants ($C_{ij}$) and elastic compliance matrix elements ($s_{ij}$) are given in **Table III**. The elastic compliances show negative values of shear components, which implies that elongation along one axis leads to contraction to the conjugate axis.

As described above, the distribution of phonon energies averaged over Brillouin zone does not show significant change with magnetic configuration; hence, we do expect that the thermodynamic quantities, which are functional of phonon density of states, might not change significantly with magnetic orientation.

## IV. RESULTS AND DISCUSSION

### A. Temperature Dependence of Phonon Spectra

We have measured the inelastic neutron scattering (INS) spectra of $YCrO_3$ and $LaCrO_3$ **(Fig. 2)** up to 550 K. The temperature range of measurements includes the magnetic transition in both the compounds ($T_N(YCrO_3)$[29] ~140 K and $T_N(LaCrO_3)$[50] ~291 K). Since both the compounds are magnetic, the INS data contain both phonon and magnetic contributions. The magnetic contribution in the INS spectrum is of two kinds, one is due to well-defined magnon excitations that dominates below the magnetic transition temperature, and other is the quasielastic scattering from paramagnetically rotating spins that dominates above the transition temperature. It is known that the magnetic contribution to the INS spectrum dominates at low momentum transfers (Q) due to the magnetic form factor. Therefore, two Q-domains were considered; i.e., high-Q (4 to 7 Å$^{-1}$) and low-Q (1 to 4 Å$^{-1}$) in order to extract the magnetic contribution in the INS data at low-Q and the phonon contribution at high-Q.

The temperature dependence of the Bose-factor corrected S(Q, E) plots of $YCrO_3$ and $LaCrO_3$ are shown in **Fig 2**. At low temperatures (up to 315 K), the low-Q data show a larger elastic line as compared to the high-Q data. We could see that the energy spectrum close to the elastic line is also very sensitive to Q, i.e. at lower Q (1 to 4 Å$^{-1}$), there is larger intensity in the low energy spectrum and at higher momentum transfer (Q~4-7 Å$^{-1}$), it reduces significantly. As noted above, the magnetic signal is more pronounced at low Q, and vanishes at higher Q, following the magnetic form factor. We speculate that this quasi-elastic scattering originates from paramagnetic spin fluctuations.

The low-Q inelastic spectrum in $YCrO_3$ shows significant intensity at energy around 20 meV at 100 K (below magnetic transition temperature $T_N$~140 K), which decays rapidly above $T_N$. This indicates that there is significant contribution by magnon excitations at about 20 meV. Further, the intensity around zero energy is contributed by paramagnetic scattering of neutrons by Cr ions. The inelastic spectrum averaged over high Q does not show any prominent change with temperature. However, in case of $LaCrO_3$, we do not observe any significant difference between the low- and high-Q data at 175 K ($T_N$~291 K) (**Fig 2)**, and so, the contribution from the magnetic excitations seems to be insignificant.

As mentioned above, the neutron inelastic scattering spectrum from powder sample contains rich information of dynamics of atoms from the entire Brillouin zone. For magnetic systems, it also contains the magnon density of states. McQueeney et al[51] have shown that the information of magnetic structure is embedded in magnon density of states and can be extracted from the powder neutron inelastic scattering data. Using the Heisenberg Hamiltonian, the authors derive the magnon dispersion relation and show that for different magnetic structure, the magnon density of states exhibits distinct features in the von Hove singularities irrespective of the magnetic exchange interaction strength. For, a three-dimension ferromagnetic (F-type) cubic perovskite, the magnon density of states shows five von-Hove singularities contributed by different high symmetry zone boundary modes. The A-type antiferromagnetic cubic perovskite will have many von-Hove singularities, while the C- type AFM structure will have only three von Hove singularity features in magnon density of states. The authors found that the G- type magnon density of states show very distinct magnon density of states and it contains only one cusp like singularity near the maximum magnon energy.

In the inelastic experiment, we measure the dynamical structure factor S(Q, E), which contains the information of both the magnon and phonon density of states. Hence, the dynamical structure factor contributed from the magnon dynamics would also exhibit distinct von-Hove singularities as described above. To probe these features in our inelastic neutron spectrum, we have further processed our data to extract the magnon structure factor $S_{mag}(Q,E)$ by subtracting the high-Q inelastic neutron spectrum from the low-Q inelastic spectrum. We obtained the $S_{mag}(Q,E)$ for $YCrO_3$ (**Fig. 3(a)**). The magnon dynamical structure factor consists of one von Hove singularity at around ~20 meV energy, which is a strong signature of the G-type AFM structure. This result supplements the previous diffraction observation and justifies our calculation with the G-type AFM structure.

Here, our aim was to describe the richness of the information contained in the INS spectrum, which can be further utilized to characterize the material properties. In order to confirm that the peak observed at around 20 meV in $S_{mag}(Q,E)$ has magnetic origin and hence should follow the magnetic form factor of associated magnetic ion, we have plotted (**Fig. 3(b)**). This Figure gives the integrated intensity in the $S_{mag}(Q,E)$ over 17-22 meV as a function of momentum transfer from 1-7 Å$^{-1}$. We have also calculated the Q dependence of magnetic form factor of $Cr^{+3}$ using analytical method[52] and compared it with the variation of the integrated peak intensity ($S_{mag}(Q,E)$) with Q. We found a very good agreement between the two, which unambiguously confirms that the peak ~20meV in $S_{mag}(Q, E)$ is contributed by magnetic excitation (magnons).

Further, the spectra derived from S(Q,E) data using Eq. (2), within the incoherent approximation in both the Q regimes, are shown in **Fig. 4**. We find that for both the compound the low-Q data show large variation in the intensity as a function of temperature. However, for the high-Q data we do not observe any significant change in the spectra except at the lowest temperature. Further, in the low-Q data of $YCrO_3$ at 100 K, which is below the magnetic transition temperature (~140 K), there is a large intensity of the low energy inelastic spectra (~20 meV) as compared to the data collected at higher temperatures. This is expected to be due to a strong magnetic signal. At 175 K, it is found that there is considerable decrease of the intensity of the low energy peaks around 20 meV indicating loss of the magnetic signal.

## B. Magnetic Ordering and Calculated Phonon Spectra

We have calculated the relaxed structure for both the compounds in the non-magnetic as well as magnetic configurations, and the results are given in **Table I**. Without the magnetism the a-lattice constant is significantly overestimated by about 5 %. We find that the magnetism substantially reduces the overestimate to about 2 %. Further, in order to see the effect of the direction of magnetic moment on the structure we performed both collinear (moment along c-axis) and non-collinear calculations (moment along a-axis) for $YCrO_3$. We find that the relaxed lattice parameters are quite similar in the two cases and differ from the experiments by about 2% (**Table I**), which is acceptable in standard DFT calculation. The calculated magnetic moment of Cr ion in both the compounds is found to be 3.0 $\mu_B$, which is due to the fact that Cr atom occurs in +3 ionic states in the compounds and all the remaining d$^3$ electron occupy the $t_{3g}$ orbital.

As noted above, the compound $YCrO_3$ shows $G_a$ type magnetic configuration where the moments are aligned along a-axis. However, to see the effect of magnetic anisotropy on phonons we have performed phonon calculation in both $G_a$ and $G_c$ magnetic configurations. The magnetic moment along c-direction can be taken care by collinear calculation, while other than the c-direction of magnetic moments can be included via noncollinear calculations only. The noncollinear calculations are much more expensive than the collinear calculations. We have performed the collinear and noncollinear calculation for YCrO$_3$ and found that change in the magnetic direction from c- to a-axis does not change the phonon spectrum significantly. Hence, we have only performed collinear calculation for LaCrO$_3$.

It may be noted that above the magnetic transition temperature the compounds become paramagnetic, where the magnetic ordering is lost which results in marginally reduced effective interaction between magnetic ions. Hence, in order to reproduce the phonon spectrum in paramagnetic phase we cannot use nonmagnetic calculation (magnetic moment of Cr= 0 $\mu_B$, which is equivalent to magnetic quenching). The calculation of phonon spectrum in paramagnetic phase can be obtained by calculating the phonon spectrum over a large set of randomly oriented spin configurations which is not a feasible option but may be in near future one can use machine learning methods to achieve the same.. However, in the present case, we can estimate the changes in the phonon spectrum between few magnetic configurations and can have a sense of change in phonon energies across the magnetic transition and this could be useful to identify modes which may show strong spin-phonon coupling.

The calculated partial density of states of various atoms in collinear and noncollinear magnetic configuration for YCrO$_3$ are shown in **Fig 5**. It may be noted that the low energy phonon spectrum below 20 meV does not show any change, while small changes are observed at higher energies (above 45 meV) that are mostly contributed by chromium and oxygen atoms. This can be understood from the fact that the magnetic interaction between chromium atoms are mediated by oxygens and change in the direction and magnitude of magnetic moment will affect the Cr-O bond length and O-Cr-O bond angle, and hence the phonons. It has been reported that the Raman modes[33] in YCrO$_3$ around ~ 560 cm$^{-1}$ (~70 meV) show significant changes across the magnetic transition temperature, and therefore are expected to be influenced by magnetic interactions, which is consistent with the above ab-initio results.

The phonon density of states of LaCrO$_3$ and YCrO$_3$ differ significantly (**Fig 5)** below 50 meV, which are dominated by Y and La dynamics. The differences are mainly attributed to difference in various bond lengths and strengths in both the compounds. However, the high energy phonon modes above 50 meV, which mainly arise from Cr-O stretching, do not change much since in the both the compound the Cr-O bond length is nearly same.

## C. Experimental and Calculated Phonon Spectra

We compare in **Fig. 6** the inelastic neutron scattering measurements at T=310 K with phonon calculations that includes the magnetic interaction (spin polarised phonon calculation). The phonon density of states obtained from measurements is neutron weighted. Hence, in order to compare with the data we have done the neutron weighting of the calculated phonon data using equation (2). The calculated phonon density of states for YCrO$_3$ shows good agreement with the measurements. We observe slightly underestimated phonon energies with respect to measurements. That could be understood better in terms of overestimated lattice parameters or under-binding effect of GGA exchange correlation function in DFT approach. In case of LaCrO$_3$, the calculated phonon density of states is found to be in good agreement with the neutron data below 55 meV, while there is significant underestimation of high energy phonon density of states. The significant underestimation of high energy modes might be attributed to strong correlation effect of La-f electrons, which is not very well accounted in DFT methods. However, qualitatively overall spectrum matches very well with the measurements.

Further, we have shown (**Fig. 6**) the contribution of individual atoms to the neutron inelastic scattering spectrum. We found that for both the compounds the contribution from oxygen atoms is dominated across the entire spectral range. The Cr and R (=Y, La) atomic dynamics are significantly contributed in the spectrum below 70 meV and 30 meV respectively.

### D. Magnetic Ordering and Calculated Zone-Centre Phonon Modes

The phonon density of states is an integrated spectrum of phonons in the entire Brillouin zone. To give a sense of quantitative changes in specific phonon modes due to collinear and non-collinear magnetic configuration in YCrO$_3$, we have shown the calculated zone-center phonon frequencies in **Fig 7** and **Table II** in the G-type magnetic configuration with magnetic moments along the a- and c-axis. One may note that a few zone-center phonon modes show significant change with magnetic configuration, indicating that these modes exhibit strong spin-phonon coupling. We find that most of the zone-center modes in YCrO$_3$ above 45 meV show significantly change in phonon energies (~ 5%) with magnetic ordering along the a- and c-axis. This also reflects strongly in the density of states calculations shown in Fig. 5. Further, we have shown the calculated zone-center phonon modes of LaCrO$_3$ in Table II. We may observe that most of the zone-center phonon modes in LaCrO$_3$ are lower in energy than that in YCrO$_3$, which might be due to larger La mass and weaker bonding between La and O. Experimental data of the zone-center modes for YCrO$_3$ and LaCrO$_3$ available from Refs [26,53] and [54,55] respectively, are in fair agreement with our ab-initio calculations.

In **Fig. 8**, we have shown the eigenvectors of some of the representative modes in YCrO$_3$, which show significant change in phonon energy with magnetic configurations. The most striking feature of these modes is that these modes are mainly dominated by oxygen dynamics. The dynamics of oxygen will cause variations in the O-Cr bond and O-Cr-O bond angle, which are critical for antiferromagnetic interaction. Hence, different types of magnetic configurations result in difference in strength of the force constants for these bonds and bond angles, which in turn reflect in changes in the phonon frequencies. As the stretching and bending modes in these compounds occur at high energies, we expect that significant changes in the high energy Raman and IR modes would be observed in presence of strong magnetic field.

**E. Thermal Expansion Behavior**

Neutron powder diffraction has been used to study the high temperature behaviour of RCrO$_3$ (R = Y, La). An orthorhombic to a rhombohedral structural phase transition has been reported[56] in LaCrO$_3$ at about 533 K, while in YCrO$_3$ measurements show[57] no transition upto 1200 K. The calculated anisotropic Grüneisen parameters averaged over Brillouin zone are shown in **Fig. 9,** which show large anisotropy at low energy below 10 meV in both the compounds. The Grüneisen parameter $\Gamma_a$, below 10 meV shows negative values in both the compounds. Moreover, LaCrO$_3$ shows very large negative values for $\Gamma_a$. The calculated elastic constants and elastic compliances in YCrO$_3$ and LaCrO$_3$ are given in TABLE III. The calculated elastic constant $C_{22}$ in YCrO$_3$ shows significantly large change with change in the direction of the magnetic moment of Cr atoms between the a- and c-axis.

To get a numerical sense of the dependence of the Grüneisen parameters for YCrO$_3$ and LaCrO$_3$ on the linear thermal expansion coefficients, we can re-express the Eq. (3) using the elastic compliance values (TABLE III) as follows:

YCrO$_3$ with magnetic moment along the c- axis:

$$\alpha_a(T) = \frac{1}{V_0}\sum_{q,j} C_{l'}(q,j,T)\,[4.4\Gamma_a - 1.2\Gamma_b - 1.7\Gamma_c] \qquad (6)$$

$$\alpha_b(T) = \frac{1}{V_0}\sum_{q,j} C_{l'}(q,j,T)\,[-1.2\Gamma_a + 4.4\Gamma_b - 1.1\Gamma_c] \qquad (7)$$

$$\alpha_c(T) = \frac{1}{V_0}\sum_{q,j} C_{l'}(q,j,T)\,[-1.7\Gamma_a - 1.1\Gamma_b + 4.7\Gamma_c] \qquad (8)$$

YCrO$_3$ with magnetic moment along the a- axis:

$$\alpha_a(T) = \frac{1}{V_0}\sum_{q,j} C_{l'}(q,j,T)\,[4.7\Gamma_a - 1.1\Gamma_b - 2.1\Gamma_c] \qquad (9)$$

$$\alpha_b(T) = \frac{1}{V_0}\sum_{q,j} C_{l'}(q,j,T)\,[-1.1\Gamma_a + 3.9\Gamma_b - 0.9\Gamma_c] \qquad (10)$$

$$\alpha_c(T) = \frac{1}{V_0}\sum_{q,j} C_{l'}(q,j,T)\,[-2.1\Gamma_a - 0.9\Gamma_b + 4.9\Gamma_c] \qquad (11)$$

and
LaCrO$_3$ with magnetic moment along the c-axis:

$$\alpha_a(T) = \frac{1}{V_0}\Sigma_{q,j} C_{l'}(q,j,T)\,[5.7\Gamma_a - 1.0\Gamma_b - 2.1\Gamma_c] \qquad (12)$$

$$\alpha_b(T) = \frac{1}{V_0}\Sigma_{q,j} C_{l'}(q,j,T)\,[-1.0\Gamma_a + 4.2\Gamma_b - 1.3\Gamma_c] \qquad (13)$$

$$\alpha_c(T) = \frac{1}{V_0}\Sigma_{q,j} C_{l'}(q,j,T)\,[-2.1\Gamma_a - 1.3\Gamma_b + 4.9\Gamma_c] \qquad (14)$$

One may note that if the Grüneisen parameters were isotropic ($\Gamma_a = \Gamma_b = \Gamma_c$) in these compounds, they would have led to positive thermal expansion along all the three directions. The calculated linear thermal expansion coefficients for YCrO$_3$ and LaCrO$_3$ as a function of temperature are shown in **Fig 10**. These results show both positive and negative linear expansion, which can be attributed to anisotropy in the Grüneisen parameters, particularly since the anisotropy in the elasticity is not very large.

There are two calculations for YCrO$_3$ for the moments along the a- and c-axis respectively, which differ only at low temperatures. Above 500 K, the values of all the three linear thermal expansion coefficients are nearly the same for the two moment directions. However, there is a small NTE along the a-axis below 200 K for the case of the moment along the a-axis. The difference in the expansion behavior in the two magnetic ordering cases at low temperature arises due to the difference in the anisotropic Grüneisen parameters values (**Fig. 9**) for low-energy phonon modes in both the calculations. For the case of the moment along a-axis, $\Gamma_c$ for low-energy modes are negative, $\Gamma_a$ values are slight positive, while $\Gamma_b$ are large positive. It is interesting to note that the three terms in Eq. (9) are competing. The large positive value of $\Gamma_b$ and negative S$_{12}$ lead to the net negative thermal expansion (NTE) along the a-axis. For the case of the moment along the c-axis, all the anisotropic Grüneisen parameters have large positive values. The comparison of the results of Eq. (6) and (9) shows that large positive Grüneisen parameters values leads to positive $\alpha_a(T)$ for the calculations performed with the moment along the c-axis.

Our calculations show that LaCrO$_3$ exhibits NTE along the a-axis up to high temperatures. In the case of LaCrO$_3$, $\Gamma_a$ exhibits negative values while the other two Grüneisen parameters exhibit positive values; hence, in Eq. (12), all the three terms lead to the NTE along the a-axis. We find that in Eqs. (13) and (14) the combination of negative compliance values (S$_{21}$ and S$_{31}$) as well as negative $\Gamma_a$ values results in positive $\alpha_b(T)$ and $\alpha_c(T)$ values. The magnitude of NTE along the a-axis is large in LaCrO$_3$ (~ -10×10$^{-6}$ K$^{-1}$ at 100 K), while YCrO$_3$ shows positive expansion along all the three axes (except for a small NTE along the a-axis at low temperatures for the case of the moment along the a-axis). The differences in the sign of the linear thermal expansion coefficient $\alpha_a$ in the two compounds is attributed to the larger negative $\Gamma_a$ in LaCrO$_3$. The linear thermal expansion coefficients along the b- and c-axis in both the compounds are positive and comparable. The volume expansion coefficient in both the compounds is positive at all temperatures.

The calculated linear thermal expansion coefficients are used to calculate the lattice parameters and volume as a function of temperature. In **Fig 11,** we have compared the calculated values with the available experimental data[56,57] on RCrO$_3$ (R=Y, La). The experimental measurements are not available in the NTE region at low temperature in YCrO$_3$. The experimental observation of high temperature thermal expansion behavior of YCrO$_3$ is well reproduced (**Fig 11**) with our calculation. As noted above, the overall phonon spectrum in YCrO$_3$ is not much sensitive to the orientation of the magnetic moments. Here also we find that calculated thermal expansion behavior in YCrO$_3$ using both the magnetic configurations (**Fig 10 and 11**) are nearly same. Both the calculations show negative thermal expansion (NTE) in YCrO$_3$ along the a-axis. The anisotropic experimental data of thermal expansion in YCrO$_3$ matches very well with our calculations, while for LaCrO$_3$ we have only experimental data of volume which also match well with the calculations.

To investigate the specific anharmonic phonon energies responsible for the NTE along the a-axis in these compounds, we have calculated the contribution to the linear thermal expansion coefficients **(Fig 12)** from phonons of various energies averaged over the Brillouin zone at 300 K in YCrO$_3$ and LaCrO$_3$ respectively. We find that phonons of energy below 20 meV (**Fig 9**) are largely behind the NTE along the a-axis. Below 20 meV, the calculated phonon density of states of LaCrO$_3$ is significantly larger in comparison to that of YCrO$_3$, resulting in larger contribution to the NTE in LaCrO$_3$ along the a-axis.

It is important to mention here that a large number of phonons in the entire Brillouin zone contribute[47,58,59] to the thermal expansion behaviour. However, to understand the anisotropic expansion behavior and large NTE along the a- axis in LaCrO$_3$, we have identified a few representative low-energy phonon modes which contribute negative expansion along the a-axis while positive expansion in other directions. We have selected the modes at the zone-center and zone-boundary since they are easier to visualize compared to the modes at general wave-vectors in the middle of the Brillouin zone. These selected modes have been plotted in **Fig 13** and involve displacements of the oxygen and La atoms. The La displacements are larger in the modes of 13.9, 15.6 and 15.7 meV than that in the 19.7 meV mode, which is consistent with the partial density of states (**Fig. 5**). The oxygen movements mainly distort the CrO$_6$ octahedra, along with small overall librations. Among these modes, the largest contraction along the a-axis is produced by the mode at 13.9 meV at (0 1/2 0), which involves large displacements of La and O atoms along the b-axis. We need to caution that this may be an oversimplified picture, but it is illustrative of the NTE behaviour.

## V. CONCLUSIONS

In this article, we have presented the ab-initio lattice dynamics calculations and inelastic neutron scattering measurements of YCrO$_3$ and LaCrO$_3$. The scattering with low momentum transfer (Q=1-4 Å$^{-1}$) is significantly dominated by magnetic scattering and shows significant changes across the magnetic transition temperature, while the phonon processes are dominated at high momentum transfer regime (Q=4-7 Å$^{-1}$) that remain essentially invariant with temperature. The calculated zone-center phonon frequencies in YCrO$_3$ compound show significant change with different magnetic configurations, which suggests strong spin-phonon coupling. Both the compounds exhibit strong anisotropic thermal expansion behavior along with negative thermal expansion along the a-axis of the orthorhombic structure. The negative thermal expansion along the a-axis in LaCrO$_3$ is found to be due to the low-energy anharmonic modes involving La vibrations and distortions of the CrO$_6$ octahedra.

**ACKNOWLEDGEMENTS**

The use of ANUPAM super-computing facility at BARC is acknowledged. SLC thanks the Indian National Science Academy for award of an INSA Senior Scientist position.

TABLE I. The calculated and experimental structures[29,30] in the orthorhombic phase (space group Pnma) of RCrO$_3$ (R=Y, La). Non-collinear and Collinear correspond to the ab-initio calculations performed with magnetic moment on Cr atoms along a- and c-axis respectively. The experimental values of the magnetic moment for YCrO$_3$ and LaCrO$_3$ are from Ref [29] and [50] respectively.

|  | LaCrO$_3$ | | | | | YCrO$_3$ | | | | |
|---|---|---|---|---|---|---|---|---|---|---|
|  | a (Å) | b (Å) | c (Å) | V (Å$^3$) | Magnetic moment ($\mu_B$) | a (Å) | b (Å) | c (Å) | V (Å$^3$) | Magnetic moment ($\mu_B$) |
| **Expt.** | 5.4813 | 7.7611 | 5.5181 | 234.81 | 3.0 | 5.524 | 7.529 | 5.238 | 217.90 | 2.66 |
| **GGA (Non-collinear)** |  |  |  |  |  | 5.580 | 7.592 | 5.274 | 223.4 | 2.7 |
| **GGA (Collinear)** | 5.611 | 7.890 | 5.567 | 246.4 | 3.0 | 5.617 | 7.674 | 5.323 | 229.4 | 3.0 |
| **GGA (Non-magnetic)** | 5.764 | 7.632 | 5.515 | 242.6 |  | 5.761 | 7.420 | 5.238 | 223.9 |  |

TABLE II: Calculated energies (in meV units) of the zone-center phonon modes of $YCrO_3$. Experimental data of $YCrO_3$ and $LaCrO_3$ are from Refs [26,53] and [54,55] respectively. Moment_a and Moment_c correspond to the ab-initio calculations performed with magnetic moment on Cr atoms along a- and c-axis respectively.

|  | YCrO$_3$ | | | |  | LaCrO$_3$ | |
| --- | --- | --- | --- | --- | --- | --- | --- |
|  | Expt | Moment_a | Moment_c | %change |  | Expt | Moment_c |
| Ag | 19.1 | 17.45141 | 17.70182 | -1.41455 | Ag |  | 10.9 |
| Ag | 23.2 | 21.2 | 20.51555 | 3.33623 | Ag | 12.9 | 15.5 |
| Ag | 35.4 | 33.43646 | 32.99088 | 1.35062 | Ag | 21.7 | 23.0 |
| Ag | 43.1 | 40.7901 | 40.14101 | 1.61703 | Ag | 31.8 | 32.1 |
| Ag | 53.5 | 49.60666 | 50.61957 | -2.00104 | Ag | 34.1 | 37.4 |
| Ag | 61.9 | 57.62502 | 57.38572 | 0.41701 | Ag | 54.5 | 50.4 |
| Ag | 71.3 | 66.28013 | 65.2766 | 1.53735 | Ag |  | 65.1 |
| Au |  | 12.31029 | 12.6514 | -2.69618 | Au |  | 10.5 |
| Au |  | 22.63872 | 22.81159 | -0.75781 | Au |  | 22.1 |
| Au |  | 29.84509 | 30.43793 | -1.94772 | Au |  | 30.7 |
| Au |  | 34.87627 | 34.20633 | 1.95852 | Au |  | 34.5 |
| Au |  | 43.16414 | 44.4295 | -2.84802 | Au |  | 43.8 |
| Au |  | 49.32082 | 52.11648 | -5.36424 | Au |  | 46.1 |
| Au |  | 61.83999 | 64.5964 | -4.26712 | Au |  | 64.5 |
| Au |  | 64.20698 | 66.53099 | -3.49312 | Au |  | 65.5 |
| B1g |  | 21.97883 | 19.73271 | 11.38272 | B1g |  | 18.8 |
| B1g | 33.7 | 31.62679 | 30.23511 | 4.60286 | B1g |  | 19.6 |
| B1g | 51.3 | 47.42286 | 47.18901 | 0.49555 | B1g |  | 40.4 |
| B1g | 69.7 | 65.26916 | 63.87435 | 2.18368 | B1g |  | 64.6 |
| B1g |  | 77.95521 | 76.20505 | 2.29663 | B1g | 89.0 | 74.7 |
| B1u |  | 19.81529 | 20.33675 | -2.56411 | B1u |  | 15.5 |
| B1u |  | 24.00402 | 23.91255 | 0.3825 | B1u | 24.4 | 20.3 |
| B1u |  | 38.98252 | 38.9803 | 0.0057 | B1u |  | 32.5 |
| B1u |  | 40.00124 | 40.31292 | -0.77315 | B1u |  | 36.5 |
| B1u |  | 44.49959 | 45.03511 | -1.18913 | B1u |  | 43.8 |
| B1u |  | 50.73268 | 51.74265 | -1.95191 | B1u |  | 47.5 |
| B1u |  | 55.46418 | 56.54616 | -1.91344 | B1u | 55.3 | 48.9 |
| B1u |  | 60.80826 | 62.50683 | -2.71742 | B1u | 71.1 | 52.4 |
| B1u |  | 65.33047 | 68.40652 | -4.49672 | B1u | 83.1 | 66.6 |
| B2g |  | 16.85527 | 17.17695 | -1.87272 | B2g | 15.5 | 12.5 |
| B2g | 27.7 | 25.55497 | 25.69943 | -0.56209 | B2g | 18.7 | 17.4 |
| B2g | 39.8 | 37.41232 | 37.86516 | -1.19593 | B2g |  | 22.8 |
| B2g |  | 45.44167 | 44.52842 | 2.05094 | B2g | 43.4 | 40.5 |
| B2g | 63.2 | 57.96401 | 58.50208 | -0.91974 | B2g | 50.4 | 47.9 |
| B2g |  | 63.05613 | 64.01392 | -1.49623 | B2g |  | 58.2 |
| B2g |  | 76.64764 | 78.62161 | -2.51072 | B2g |  | 77.8 |
| B2u |  | 22.0818 | 20.66271 | 6.8679 | B2u | 17.1 | 19.4 |
| B2u |  | 24.54895 | 25.05331 | -2.01312 | B2u |  | 23.0 |
| B2u |  | 40.95023 | 41.09372 | -0.34916 | B2u | 33.0 | 40.3 |
| B2u |  | 42.47898 | 44.14774 | -3.77994 | B2u | 41.2 | 43.6 |
| B2u |  | 51.30122 | 53.86873 | -4.76625 | B2u | 44.3 | 47.8 |
| B2u |  | 59.93587 | 63.39751 | -5.46022 | B2u |  | 64.0 |
| B2u |  | 65.36872 | 66.51227 | -1.71931 | B2u |  | 65.7 |
| B3g | 21.8 | 19.97432 | 18.89256 | 5.72582 | B3g | 17.9 | 16.3 |
| B3g |  | 38.49226 | 37.72471 | 2.0346 | B3g |  | 35.4 |
| B3g | 60.9 | 55.41178 | 55.94586 | -0.95464 | B3g | 52.7 | 49.4 |
| B3g |  | 65.9585 | 64.43922 | 2.3577 | B3g |  | 64.7 |
| B3g |  | 81.42026 | 76.46348 | 6.48254 | B3g | 73.2 | 76.3 |
| B3u |  | 18.03823 | 18.32628 | -1.57183 | B3u |  | 15.4 |
| B3u |  | 27.54482 | 27.99148 | -1.59568 | B3u | 20.6 | 20.4 |
| B3u |  | 33.25443 | 33.80634 | -1.63255 | B3u | 30.0 | 31.6 |
| B3u |  | 38.85506 | 38.9792 | -0.31848 | B3u |  | 35.7 |
| B3u |  | 44.86611 | 45.25838 | -0.86671 | B3u | 47.3 | 39.1 |
| B3u |  | 47.3538 | 48.54302 | -2.44982 | B3u | 52.7 | 43.4 |
| B3u |  | 57.07838 | 58.55945 | -2.52917 | B3u | 58.4 | 52.3 |
| B3u |  | 62.63429 | 65.42302 | -4.26262 | B3u | 61.2 | 62.8 |
| B3u |  | 66.33035 | 67.33851 | -1.49714 | B3u | 75.2 | 64.5 |

TABLE III: The calculated elastic constants and elastic compliances $YCrO_3$ and $LaCrO_3$. $C_{ij}$ and $S_{ij}$ are in the units of GPa and $10^{-3}$ $GPa^{-1}$ respectively. Moment_a and Moment_c correspond to the ab-initio calculations performed with the magnetic moment on the Cr atoms along a- and c-axis respectively.

| ij($C_{ij}$/$S_{ij}$) | $C_{ij}$(in GPa) | | | $S_{ij}$(in $10^{-3}$ GPa) | | |
|---|---|---|---|---|---|---|
| | $YCrO_3$ | | $LaCrO_3$ | $YCrO_3$ | | $LaCrO_3$ |
| | Moment_a | Moment_c | Moment_c | Moment_a | Moment_c | Moment_c |
| 11 | 313.5 | 315.8 | 238.0 | 4.7 | 4.4 | 5.7 |
| 22 | 320.0 | 290.0 | 294.4 | 3.9 | 4.4 | 4.2 |
| 33 | 292.7 | 289.4 | 287.4 | 4.9 | 4.7 | 4.9 |
| 12 | 128.4 | 121.9 | 92.8 | -1.1 | -1.2 | -1.0 |
| 13 | 160.0 | 144.1 | 126.0 | -2.1 | -1.7 | -2.1 |
| 23 | 114.2 | 110.1 | 115.6 | -0.9 | -1.1 | -1.3 |
| 44 | 78.4 | 80.6 | 84.2 | 12.8 | 12.4 | 11.9 |
| 55 | 100.6 | 98.6 | 96.0 | 9.9 | 10.1 | 10.4 |
| 66 | 107.0 | 106.8 | 90.8 | 9.3 | 9.4 | 11.0 |

FIG. 1 (Color online) The structure of orthorhombic RCrO$_3$ (R=Y, La). Key: Y/La – Silver, Cr– Blue and O – Red.

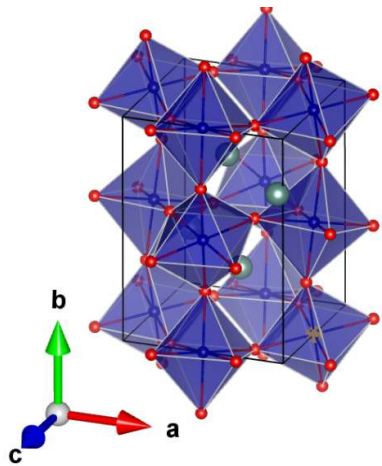

FIG. 2 (Color online) Temperature-dependent inelastic neutron scattering spectra of RCrO$_3$ (R=Y, La). Top panel: the low-Q and high-Q Bose-factor corrected S(Q, E), where both the energy loss (0 - 10 meV) and the energy gain (-100 - 0 meV) sides are shown.

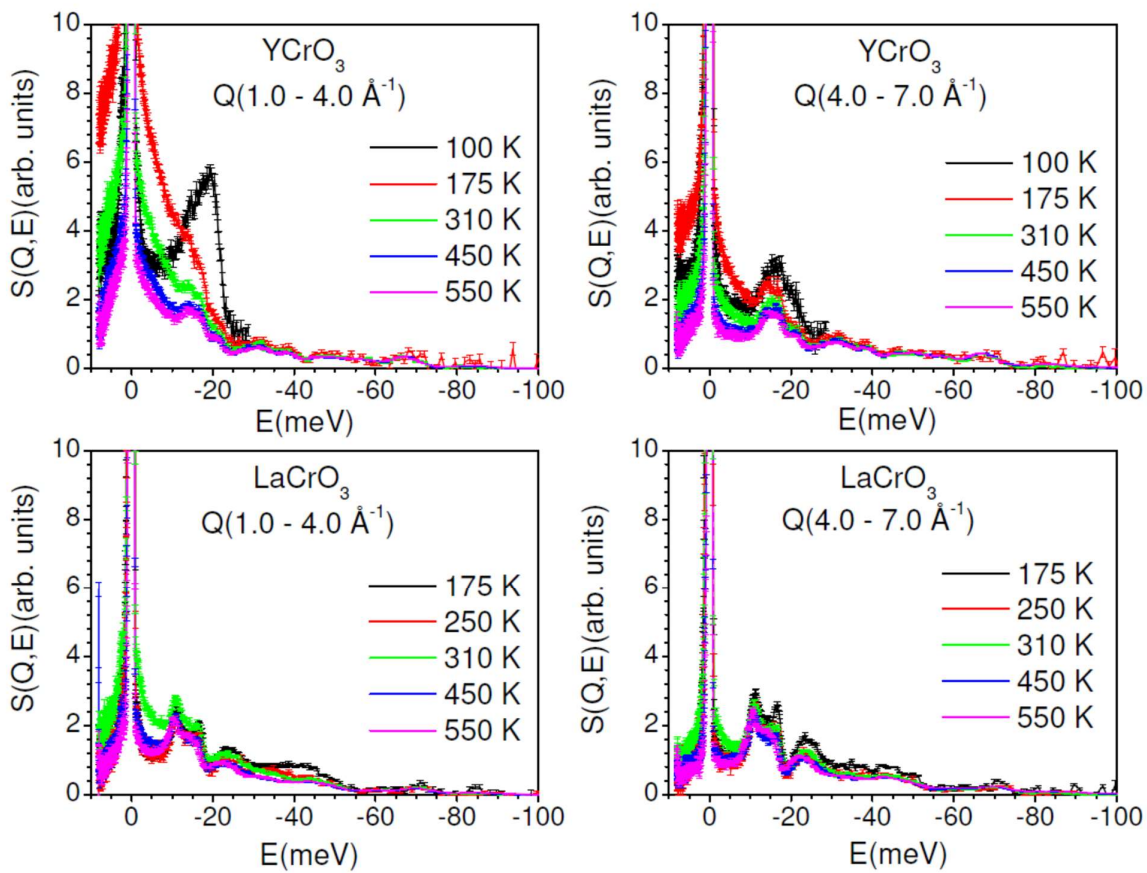

FIG. 3 (a) Extracted magnetic contributions (intensity difference between low-Q and high-Q) in neutron inelastic spectra of YCrO$_3$ at temperatures below T$_N$ from polycrystalline samples. (b) The variation of integrated intensity of the excitation for YCrO$_3$ (energy range = 17-22 meV) with wave-vector Q. The magnetic form factor for Cr$^{3+}$ is computed using the analytical relation and shown as a continuous line. The experimental intensity is scaling by a constant factor for comparison with the calculations.

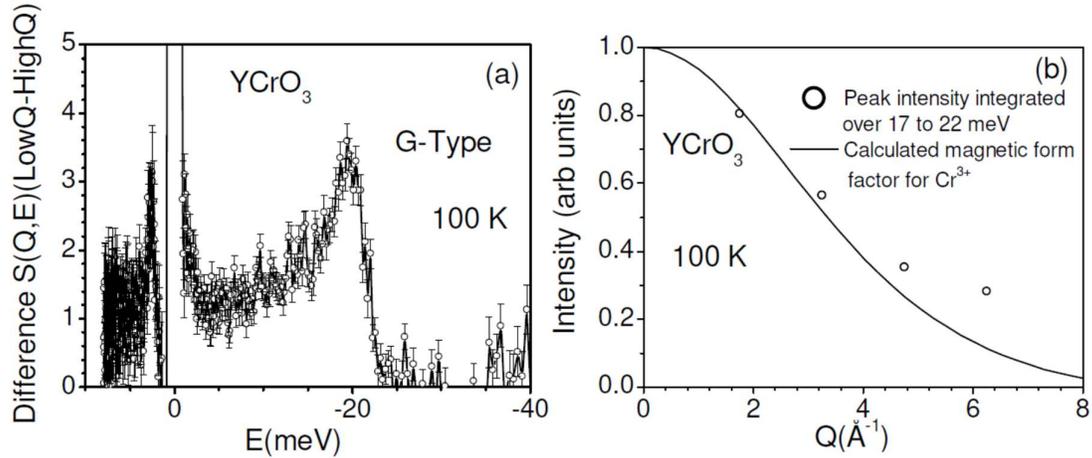

FIG. 4 (Color online) Temperature-dependent neutron inelastic scattering spectra of RCrO$_3$ (R=Y, La). The low-Q and high-Q, unity-normalized, excitation spectra, g$^{(n)}$(E), inferred from the neutron energy gain S(Q,E) data, within the incoherent approximation.

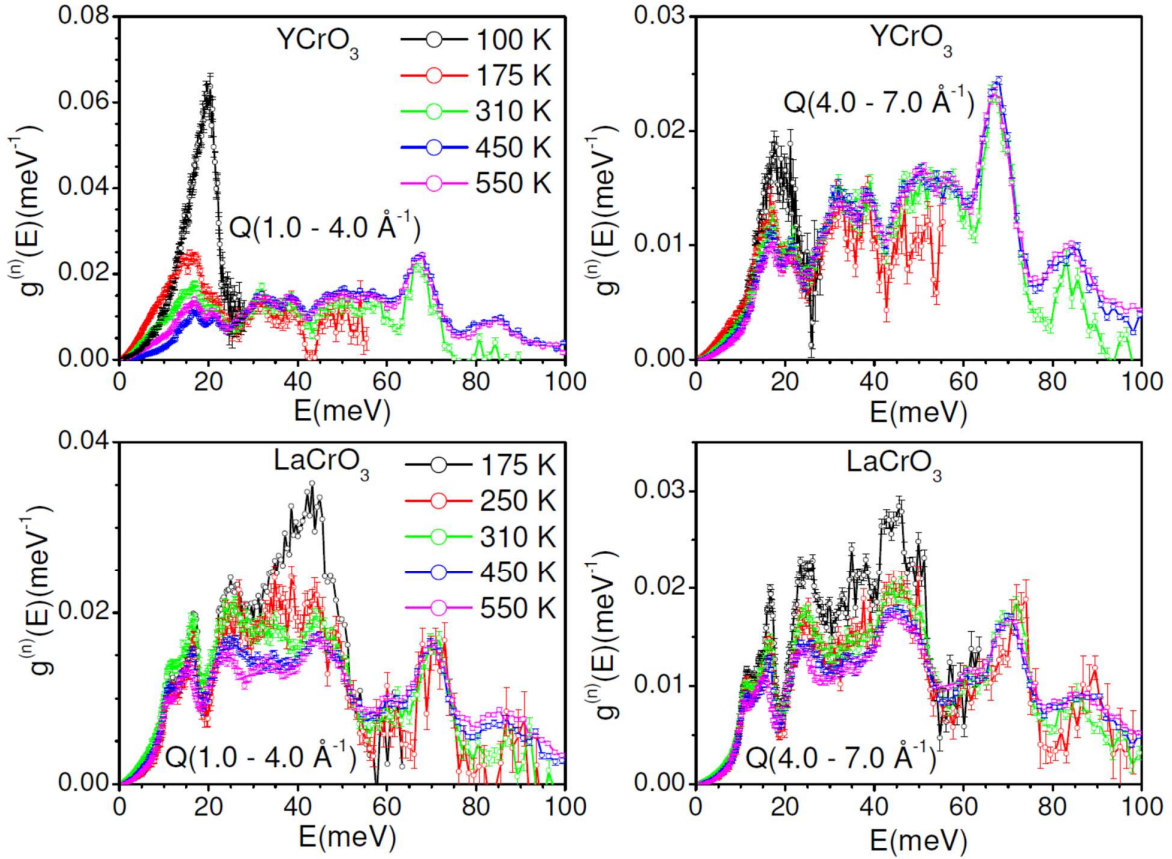

FIG. 5 (Color online) The calculated atomic partial phonon density of states (Y/La, Cr and O) in the orthorhombic phase of RCrO$_3$ (R=Y, La).

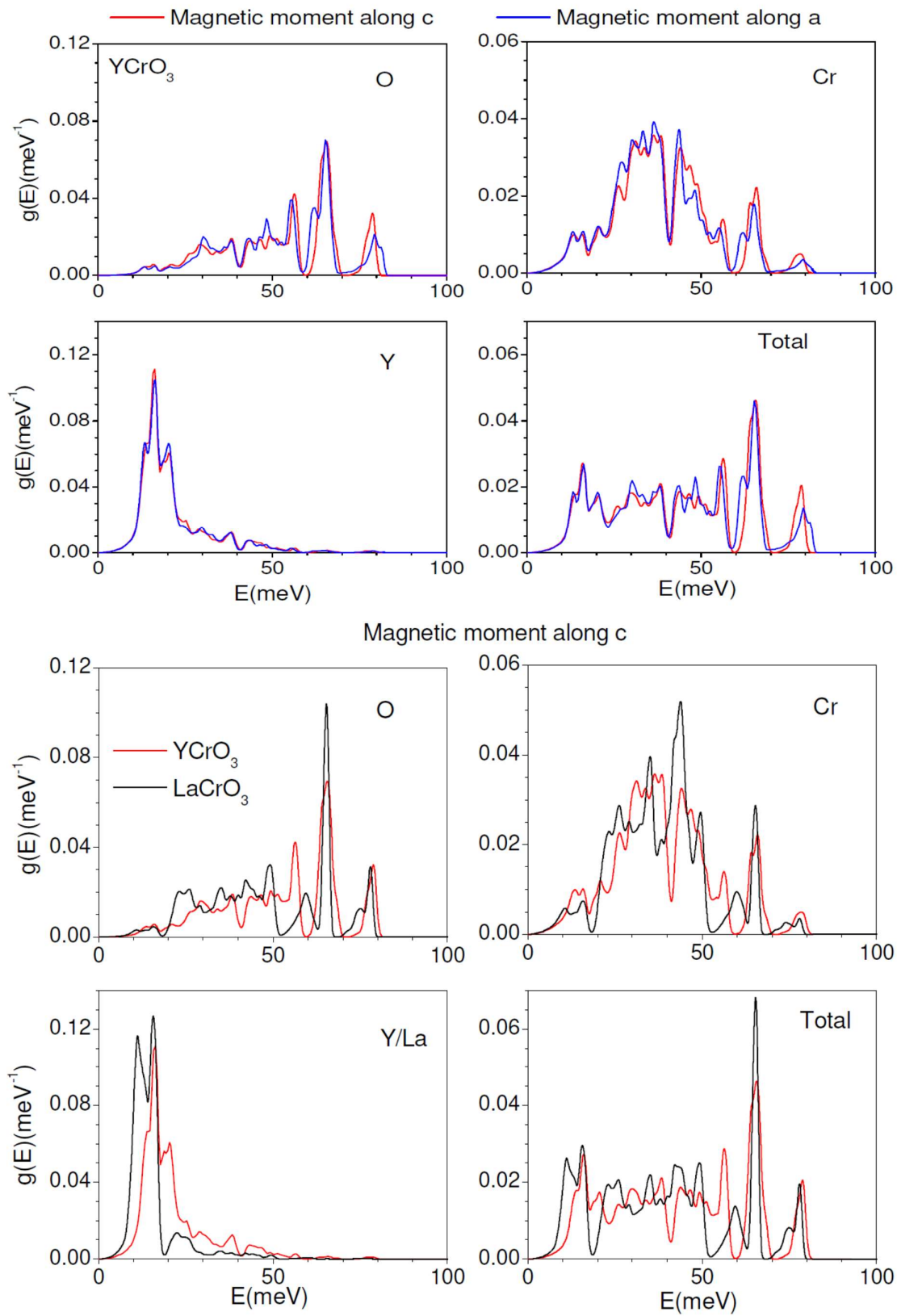

FIG. 6 (Color online) The calculated and experimental neutron inelastic scattering spectra of RCrO$_3$ (R=Y, La).

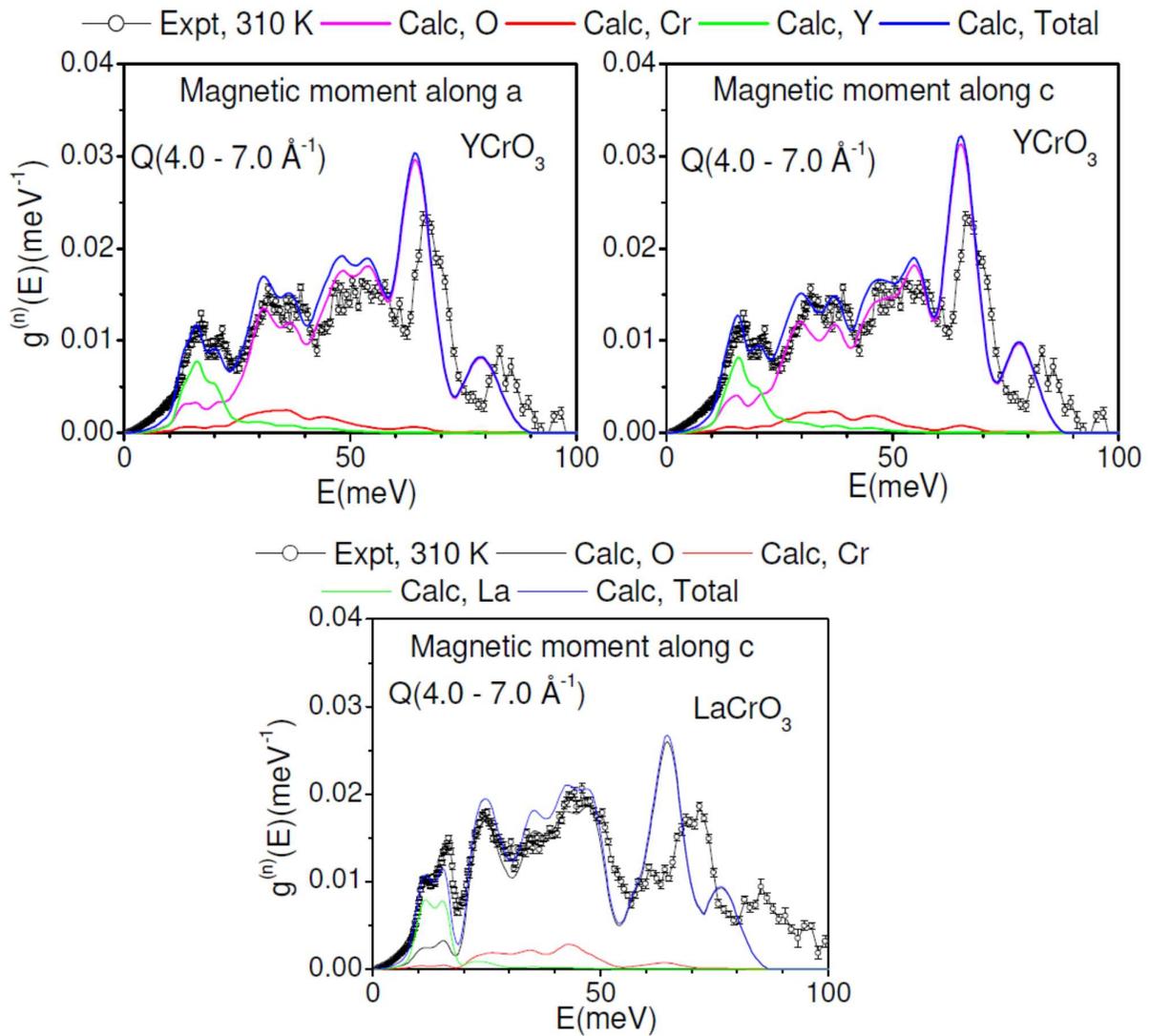

FIG. 7 (Color online) The calculated zone-center frequencies in collinear and noncollinear magnetic configurations in YCrO$_3$ and their irreducible representations.

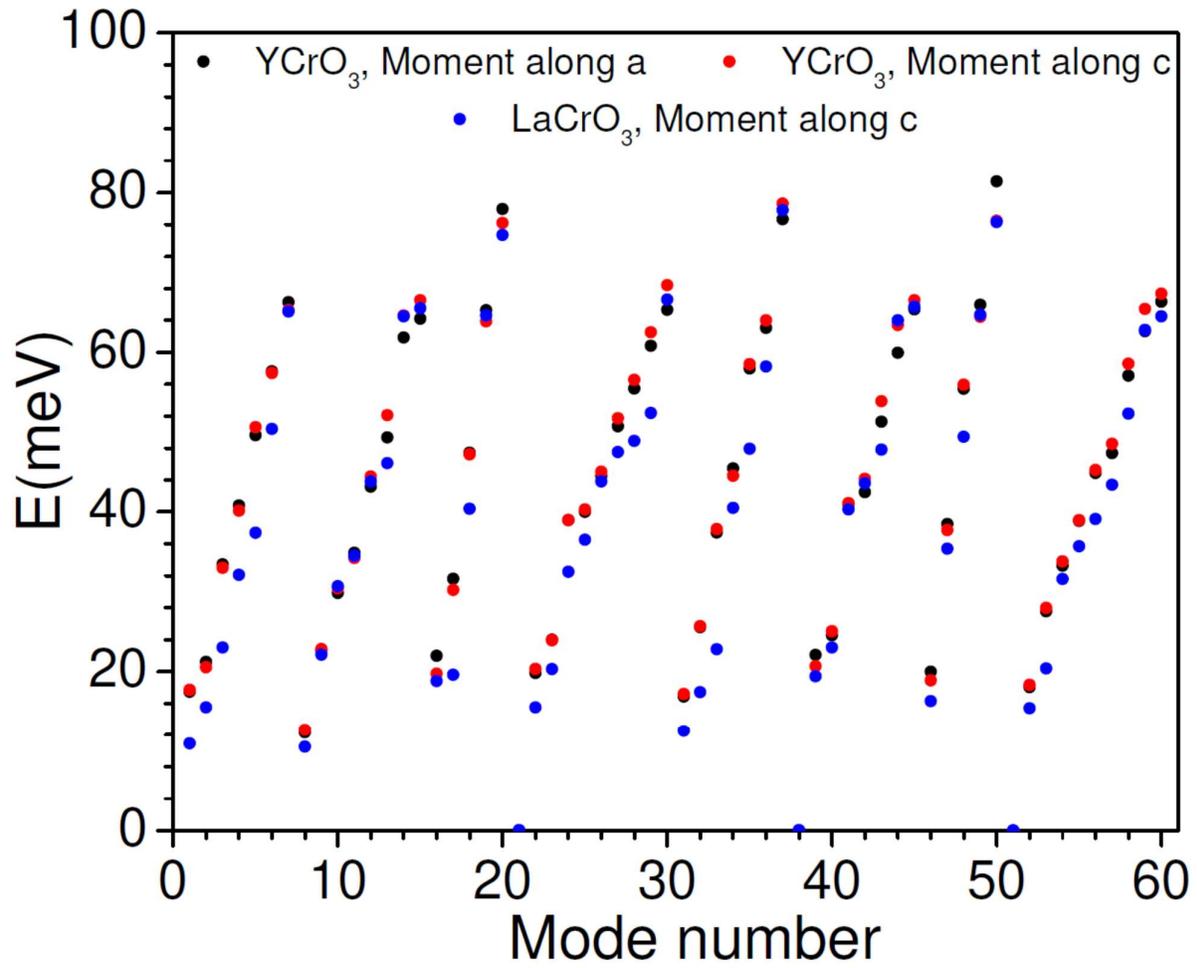

FIG 8 (Color online): The displacement pattern of representative zone-center modes that exhibit large change in phonon energy in the two magnetic configurations (the moment along a- axis and c- axis respectively) in YCrO$_3$. The numbers below each plot represent the character of the mode, and the numbers in the next line are the phonon energy in meV with the magnetic moment along a -axis and c-axis respectively and the percentage change. Key: Y – Silver, Cr– Blue and O – Red.

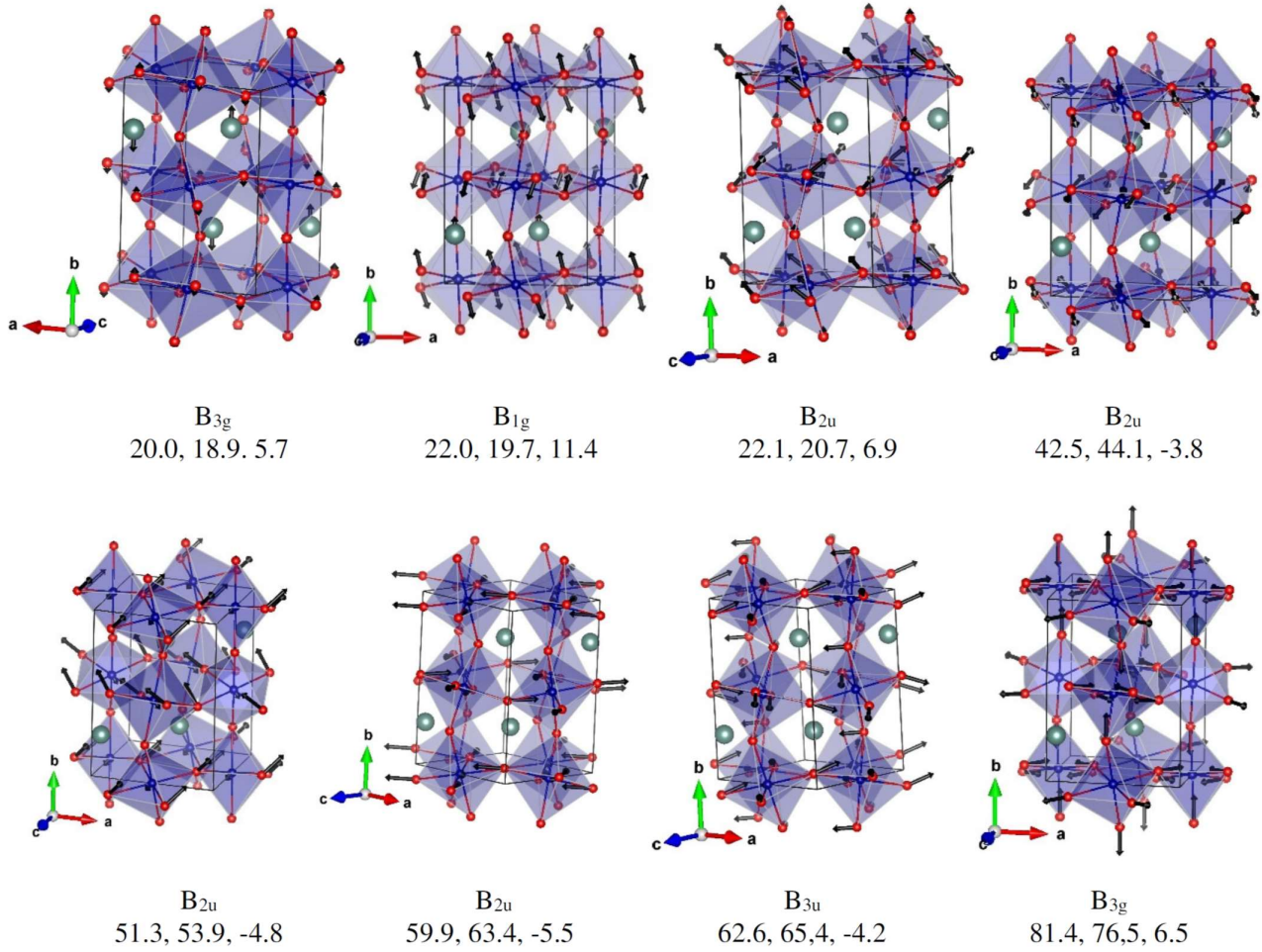

B$_{3g}$
20.0, 18.9. 5.7

B$_{1g}$
22.0, 19.7, 11.4

B$_{2u}$
22.1, 20.7, 6.9

B$_{2u}$
42.5, 44.1, -3.8

B$_{2u}$
51.3, 53.9, -4.8

B$_{2u}$
59.9, 63.4, -5.5

B$_{3u}$
62.6, 65,4, -4.2

B$_{3g}$
81.4, 76,5, 6.5

FIG 9 (Color online): Variation of the anisotropic Grüneisen parameters with phonon energy E averaged over the Brillouin zone in RCrO$_3$ (R=Y, La).

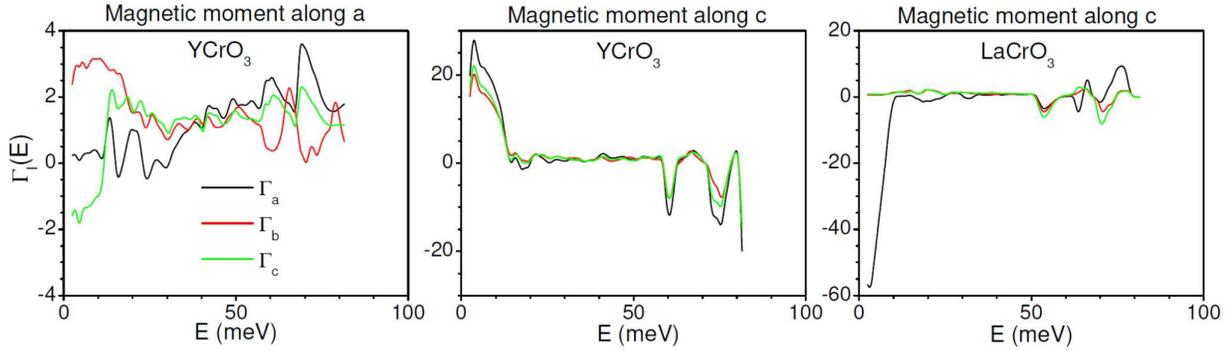

FIG 10 (Color online): Calculated anisotropic thermal expansion in RCrO$_3$ (R=Y, La). Calculations for LaCrO$_3$ is shown in the orthorhombic phase[56], which is stable up to about 533 K.

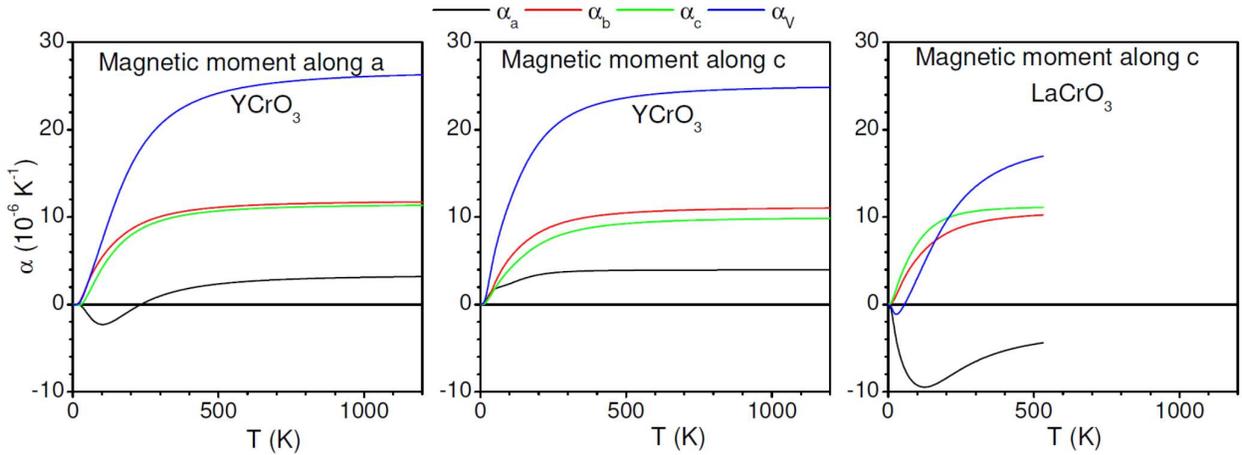

FIG 11 (Color online): Variation of $a$, b and $c$ lattice parameters, and volume (X=a, b, c, V) with temperature in RCrO$_3$ (R=Y, La). The experimental thermal expansion data[56,57] on RCrO$_3$ (R=Y, La) (solid circles) are from literature. Calculations for LaCrO$_3$ are shown in the orthorhombic phase[56], which is stable up to about 533 K.

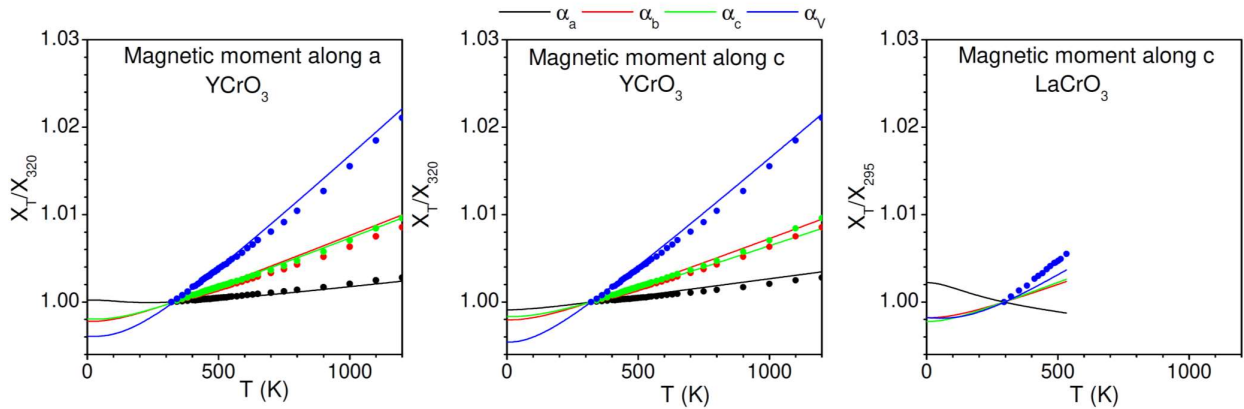

FIG 12 (Color online): The contribution to anisotropic linear thermal expansion coefficients at 300 K, from phonon modes of energy E averaged over the Brillouin zone.

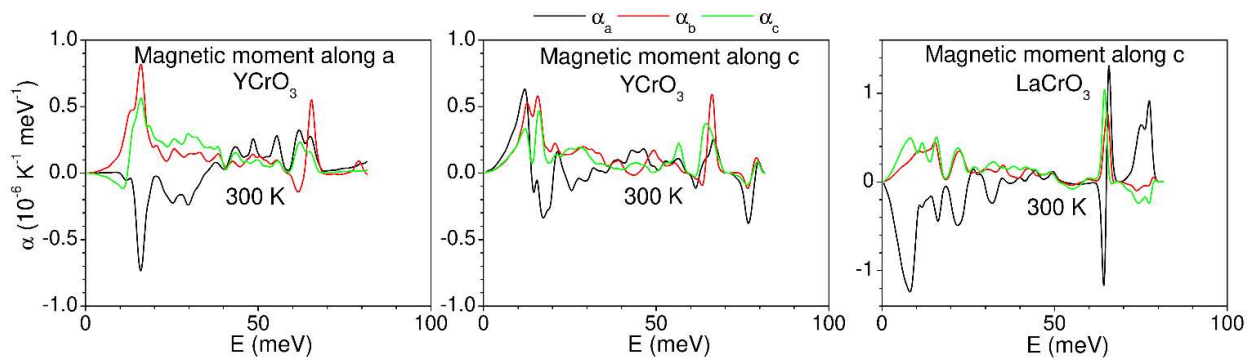

FIG 13 (Color online): The displacement pattern of representative zone-center and zone-boundary modes in LaCrO$_3$ which exhibit negative thermal expansion behavior along the a-axis. The values of the linear thermal expansion coefficients ($\alpha_a$, $\alpha_b$ and $\alpha_c$) from each mode (assuming it to be an Einstein oscillator) are given at 300 K in the units of $10^{-6}$ K$^{-1}$. Key: La – Silver, Cr – Blue and O – Red.

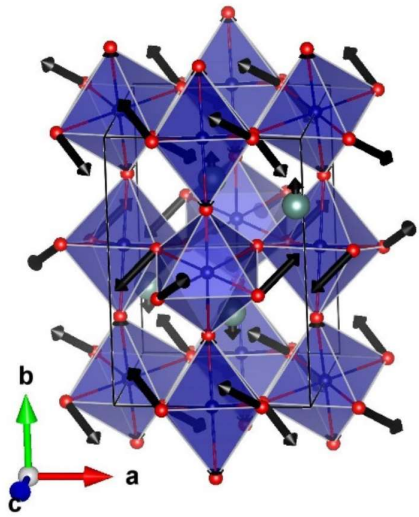

q=(0 0 0), 19.6 meV
$\Gamma_a$=-1.3, $\Gamma_b$=5.1, $\Gamma_c$=1.7
$\alpha_a$=-0.85, $\alpha_b$=1.10, $\alpha_c$=0.24

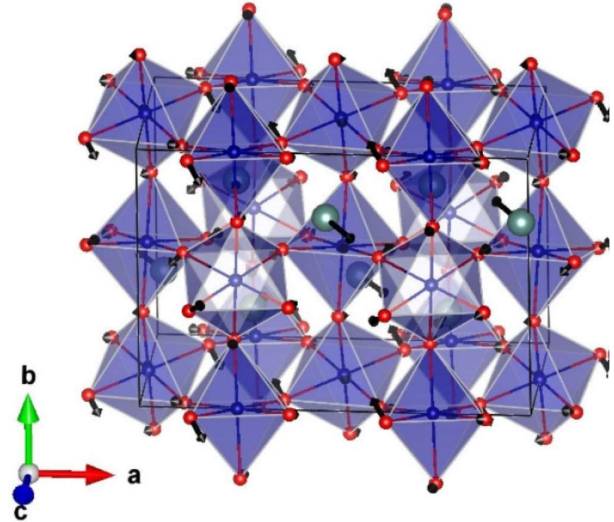

q=(0.5 0 0), 15.6 meV
$\Gamma_a$=-1.5, $\Gamma_b$=4.2, $\Gamma_c$=1.7
$\alpha_a$=-0.87, $\alpha_b$=0.91, $\alpha_c$=0.34

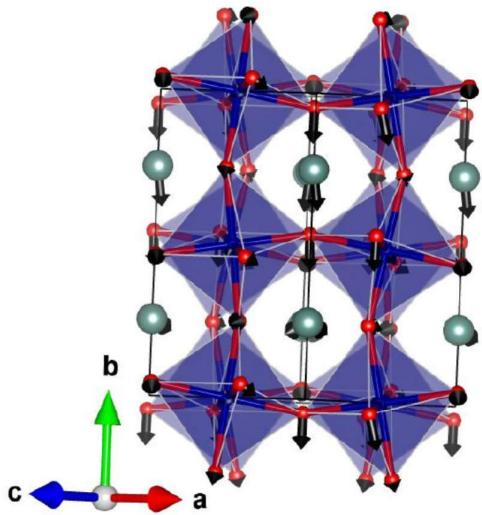

q=(0 0.5 0), 13.9 meV
$\Gamma_a$=-8.9, $\Gamma_b$=3.9, $\Gamma_c$=2.1
$\alpha_a$=-3.22, $\alpha_b$=1.22, $\alpha_c$=1.33

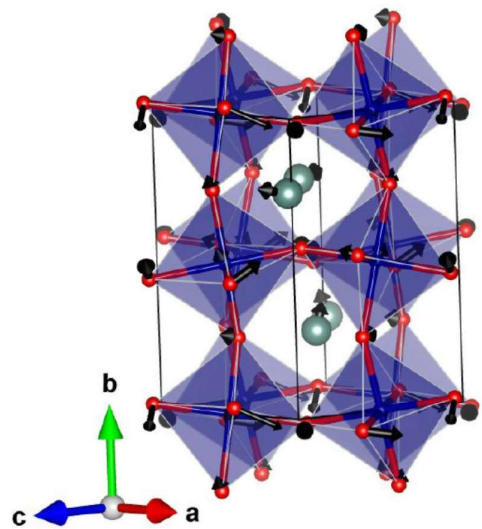

q=(0 0.5 0), 15.7 meV
$\Gamma_a$=-3.6, $\Gamma_b$=1.6, $\Gamma_c$=2.9
$\alpha_a$=-1.53, $\alpha_b$=0.35, $\alpha_c$=1.08